# Independent Ion Migration in Suspensions of Strongly Interacting Charged Colloidal Spheres

Dirk Hessinger, Martin Evers, Thomas Palberg
Universität Mainz, Inst. f. Physik, D-55099 Mainz, Germany

**Abstract**

We report on sytematic measurements of the low frequency conductivity $\sigma$ in aequous supensions of highly charged colloidal spheres. System preparation in a closed tubing system results in precisely controlled number densities of $10^{16}$ m$^{-3}$ ≤ n ≤ $10^{19}$ m$^{-3}$ (packing fractions of $10^{-7}$ ≤ $\Phi$ ≤ $10^{-2}$) and electrolyte concentrations of $10^{-7}$ ≤ c ≤ $10^{-3}$ mol l$^{-1}$. Due to long ranged Coulomb repulsion some of the systems show a pronounced fluid or crystalline order. Under deionized conditions we find $\sigma$ to depend linearly on the packing fraction with no detectable influence of the phase transitions. Further at constant packing fraction $\sigma$ increases sublinearily with increasing number of dissociable surface groups N. As a function of c the conductivity shows pronounced differences depending on the kind of electrolyte used. We propose a simple yet powerful model based on independent migration of all species present and additivity of the respective conductivity contributions. It takes account of small ion macro-ion interactions in terms of an effectivly transported charge. The model successfully describes our qualitatively complex experimental observations. It further facilitates quantitative estimates of $\sigma$ over a wide range of particle and experimental parameters.

**Introduction**

The electrokinetics of charge stabilized colloidal particles have attracted considerable theoretical and experimental interest. For a recent review see e.g. Lyklema [1]. Phenomena like electrophoresis, electroosmosis, sedimentation potential, electroviscous effects, diffusion and low as well as high frequency dielectric properties have been investigated both experimentally and theoretically as function of particle charge Z, electrolyte concentration c and particle concentration. However, for most of these experiments and even in the case of isolated particles a full electrohydrodynamic treatment of macro-ions, small ions, their interactions with each other and with the externally applied electric fields is usually necessary to describe the complexity experimental observations.

On the other side highly charged colloidal suspensions have been recognized as valuable model systems for a number of important condensed matter problems, since they tend to form fluid-like or crystalline ordered states due to the long ranged electrostatic repulsion [2, 3]. Questions addressed include stucture formation as well as dynamics or phase transition kinetics [4, 5, 6]. In contrast to electrokinetics, these aspects can often be treated using an effective pair interaction of Debye-Hückel (DH) type with all details covered by an effective macro-ion charge smaller than the bare particle charge. There exist procedures to numerically calculate this renormalized charge on the basis of the mean field description in terms of the non-linear Poisson-Boltzmann (PB) equation [7] as well as support from more sophisticated treatments [8,9] and many experimental studies.

The simplicity of such an approach is appealing, but it is not immediately clear whether it may be useful for the description of conductivity. If we recall the situation encountered in simple electrolytes, dilute aequous solutions of strong 1:1 electrolytes usually show additivity of their conductivity. The conductivity of the solution is easily calculated from the sum of the limiting molar conductivities of each ionic species at infinite dilution times their respective molar concentration. At higher concentrations Ostwald´s rule applies to correct for concentration effects on the molar conductivities; additivity nevertheless remains valid. For very dilute systems ($c < 10^{-5}$ mol l$^{-1}$) additivity also applies for mixed electrolytes. It is an interesting open question, whether this concept remains valid, if one considers a mixture of a simple 1:1 electrolyte and a highly assymetric Z:1 electrolyte, where Z may for micelles, polyelectrolytes or colloidal particles be on the order of 10 to $10^5$.

In this contribution we will as a first step try to extend the concept of additivity of conductivities as known from simple electrolytes with the aid of an effectively transported charge $Z^*_\sigma$. This explicetely includes the cases where the interaction is sufficiently strong to form an ordered suspension. The interesting question of a connection of this effectively transported charge to the numerically accessible renormalized charge [7] will be shortly addressed, too, while a comprehensive test has to be postponed to a subsequent work.

A number of complications arise in contrast to simple electrolytes. Firstly, the interaction between macro-ions and small ions has to be considered. As a consequence the distribution of the small ions is no longer uniform on the length scale of the Debye screening length but the formation of the electric double layer (EDL) around the macro-ions has to be described in terms of solutions to the PB equation. For the macro-ion this in turn leads to deviations from the Nernst formula giving a simple analytic relation between the diffusivity, the mobility in an electric field and the charge of an ion. For the small ions the mobility becomes dependent on their radial distance from the macro-ion surface either due to the „binding" potential of the central macro-ion and/or due to interactions between the small ions themselves which become important at elevated concentrations near the surface. Finally, however a high macro-ion charge may, in addition to electrostatically stabilizing the system against coagulation, cause the establishment of fluidlike or even crystalline order amongst the macro-ions in the suspension.

In experimental studies on the low frequency conductivity $\sigma$ of colloidal suspensions much attention was paid on the dependence of $\sigma$ on the particle packing fraction. In deionized samples a linear increase is observed for all packing fractions in the range of $10^{-4}$ to $2 \times 10^{-2}$ [10, 11], while for samples containing background electrolyte this is found at elevated packing fractions ($\Phi <$ ) only [12]. In all these cases, however, the conductivity was found lower than expected on the basis of the bare particle charge. Explicit studies of the charge dependence of the conductivity are rare. In a recent study of Yamanaka *et al*. a continued sublinear increase was observed with the increasing number of dissociateble surface groups N for silica particles charged *via* a chemical reaction with NaOH [13]. A similar finding was reported by us in a previous work on particles with physisorbed anionic surfactant [14]. This system, FEP(PFOA)78, will be studied again and in more detail here.

At fixed packing fraction the conductivity depends on the kind of electrolyte added. Recently Sumaru *et al.* [15] and independently Swetslot and Leyte [16] investigated the influence of adding different 1:1 electrolytes (HCl, NaCl, NaOH) and 2:1 electrolyte ($BaCl_2$), respectively. Interestingly, upon addition of neutral electrolyte, like NaCl, a pronounced nonlinearity in $\sigma(c)$ was observed which was not known from former studies at higher electrolyte concentrations. A systematic data base collected over a wide range of parameters, however, is still missing.

Further, the important issue of the form of conductivity titration curves has been addressed [17, 18, 19]. For the case of weak acids, the presence of more than one acidic species or variations in the background electrolyte, the shape of the titration curve may become fairly complex and considerable effort is necessary for its interpretation. In the case of strongly acidic surface groups an initially linear decrease in $\sigma$ with increased NaOH concentration is followed by a transient region, which crucially depends on the details of surface chemistry. Far beyond the equivalence point one observes a linear increase. For particles bearing only one kind of strongly acidic surface groups a simple extrapolation of the linear regimes enables the determination of the surface group number N. Open questions e.g. concern the slope of the linear regimes which may considerably deviate from expected values.

Theoretical explanation of experimental results was given on different levels of sophistication. Only few authors used the full electrohydrodynamic treatment of frequency dependent conductivity contributions from small ion transport by the bulk field and along particle surface via electroosmosis, particle transport via electrophoresis, alterations in the ionic distribution and polarization effects [20] (see also [1] for an overview). For a given experimental situation [17] an excellent agreement with the data can be acchieved. The major drawback here is that the applicability of results is generally restricted to isolated particles and often also to large amounts of added salt, and that considerable numerical effort is necessary to perform predictions over a wide range of parameters.

On the macroscopic level the most basic treatments consider small ions only [10]. All alterations as compared to the expected value are treated using an effective number of charges, respectively counterions, much in the spirit of the charge renormalization concept. This also holds for early treatments explicetely including particle contributions [21, 22]. More recently a number of authors used a different approach explicetely considering changes in the small ion mobility but retaining the bare charge. These approaches for instance consider changes in the number of contributing small ions through surface chemical reactions and/or association of ionic species to the particle surface [12, 19, 23] as well as radial changes in small ion conductivity contributions due to the non-uniform ion distribution around the particles *via* solutions of the linearized or non-linearized PB-equation [15, 16, 19]. James *et al*. [19] accounted for the salt concentration dependence of small ion conductivities using the theory of Robinson and Stokes [24]. Sumaru *et al*. [15] consider changes in small ion mobilities through an extension of the theory of Oshima *et al*. [25]. Zwetslot and Leyte [16] combined Jönssons theory of small ion diffusion as function of their concentration which accounts for both hard sphere and electrostatic ion-ion interactions [26] with the solution of the non-linearized PB-equation in a version modified to include surface dissociation equilibrium and with Nernst´s formula relating diffusivity to mobility. Again experimental data are well described by the theoretical approach. The slight underestimation of conductivities at very low salt concentrations probably is due to the complete neglection of any particle contribution. These works [15, 16, 19] present very satisfying theoretical treatments of low frequency conductivity experiments. Using concentration or interaction dependent small ion mobilities, they nevertheless afford significant numerical effort and have been tested only in the case of comparably large amounts of added salt where the condition of isolated particles is still safely assumed to hold.

The present study differs from previous ones in two major aspects. Firstly, in this study we will employ additivity in combination with constant bulk small ion mobilities and an effective charge $Z^*$ lower than the bare particle charge. For this purpose we extend the model of independent ion migration recently proposed by Deggelmann *et al*. [21]. To be more specific, we assume simple additivity of the conductivity of all ionic species, however in addition we propose a reservoir of $Z-Z^*$ counterions in a region close to the particle which do not contribute to the conductivity but have free exchange with the outer $Z^*$ counter ions and the added electrolyte ions. In introducing the reservoir, this model is of similar spirit as the treatment of the hydrodynamic radius in diffusion problems or the shear plane concept in electrokinetics [1]. The conceptual ease, however stems from the successful extension of the treatment of simple 1:1 electrolytes to a highly asymmetric system and consequently a notation in terms of added electrolyte per particle instead of a notation in terms of an electrolyte concentration dependent and particle specific conductivity increment. The treatment therefore has only one free parameter, namely the effectively transported charge $Z^*_\sigma$.

Secondly, carefully and extensively characterized monodisperse suspensions were conditioned using advanced preparational procedures. This allowed to considerably extend the range of accessible conditions. We thus were able to thoroughly test our model with comprehensive measurements under variation of all relevant quantities entering the model like electrolyte concentration, particle concentration, particle charge and kind of electrolyte. Electrolyte concentrations for instance were well controlled even at values as low as $10^{-6}$ mol l$^{-1}$. Under these conditions most samples except for the lowest packing fractions show a pronounced fluid or even crystalline order. Particular attention was paid to the combined packing fraction and salt concentration dependence, and the use of particles stabilized by physisorbed anionic surfactant allowed measurements in dependence on the particle charge.

The paper is organized as follows. In the following chapter we give a detailed description of our samples, the preparational procedures and the experimental techniques. We then present our experimental results. In the next chapter we introduce our conductivity model and perform some calculations of conductivities under various boundary conditions. This is followed by a detailed discussion of our data in comparison to the model. There also the issue of charge renormalization is shortly addressed.

**Sample preparation and experimental methods**

Several samples of commercially available polystyrene (PS) latex spheres, one sample of perflourinated spheres (FEP(PFOA)78), kindly provided by the group of V. Degiorgio [27] and two PS-samples synthesized and kindly provided by the group of M. Ballauff [28] were used. All PS-samples were stabilized by Sulphate surface groups stemming from the polymerisation initiator and sample PS(SDS)102 in addition carried a considerable amount of physisorbed Sodium Dodecyl Sulphate (SDS, Merck,

Germany). FEP(PFOA)78 is sythesized as tetraflouroethylene copolymer with hexafluoropropene. In addition to a very small amount of perflourinated carboxyl groups the FEP-spheres were charge-stabilized by a strongly physisorbed perfluorinated anionic surfactant perflouro-octanoate acid (PFOA) [27, 29, 14]. This surfactant is water soluble only at pH > 7. The main particle features are compiled in Table 1.

| Sample | a / nm | pK | $\mu_P$ / $m^2V^{-1}s^{-1}$ | N | $Z_{PBC}$ | $Z^*_{DH}$ | $Z^*_\sigma$ |
|---|---|---|---|---|---|---|---|
| PS109 | 54. | 0.5 | $5.0 \times 10^{-8}$ | 1200±50 | ---- | 450±20 | **510±10** |
| PS115 | 57. | 0.5 | $6.7 \times 10^{-8}$ | 3600±100 | 3300±50 | 805±50 | **730±10** |
| PS120 | 60. | 0.5 | $6.7 \times 10^{-8}$ | 3600±100 | 3300±50 | ---- | **685±10** |
| PS301 | 150. | 0.5 | $6.8 \times 10^{-8}$ | 23100±300 | 21400±100 | 2440±100 | **1850±50** |
| PS(SDS)102 | 51. | 0.5 | $4.0 \times 10^{-8}$ | ---- | ---- | ---- | **800±25** |
| PSPSS70 | 35. | 0.5 | $5.8 \times 10^{-8}$ | 11800±200 | 4580±50 | 570±50 | **790±10** |
| FEP(PFOA)78 | 39. | 4.0 | $5.2 \times 10^{-8}$ | 75 - 1400 | 75 - 590 | 72 - 380 | **110 - 372** |

Table 1: Particle data. a: nominal radius as given by the manufacturer. pK as given by the manufacturer. $\mu_P$: plateau value of particle electrophoretic mobility [31, 29]. N: surface group number from titration. $Z_{PBC}$ bare charge number calculated using the PBC program. $Z^*_{DH}$: renormalized charged calculated from the PBC-program for a deionized case using a, N and pK as shown and well as $\Phi = 10^{-4}$ and $c_B = 2 \times 10^{-7}$ mol l$^{-1}$. $Z^*_\sigma$: effectively transported charge derived from packing fraction dependent conductivities of deionized suspensions.

Samples were shipped at packing fractions $\Phi$ between 0.055 and 0.25. From these we prepared stock suspensions of approximately $\Phi = 0.01$ by dilution with destilled water. To the PS-samples mixed bed ion exchange resin (IEX) [Amberlite UP 604, Rohm & Haas, France] was added and the suspensions were left to stand with occasional stirring for some weeks. They were then filtered using Millipore 0.5 μm filters to remove dust, ion-exchange debris and coagulate regularily occurring upon first contact of suspension with IEX. A second batch of carefully cleaned IEX filled into a dialysis bag was then added to retain low ionic strength in the stock suspensions now kept under Ar-atmosphere. FEP(PFOA)78 and PS(SDS)102 were extensively dialysed against destilled and decarbonized water with IEX added to the reservoir

All further sample preparation and the measurements were performed in a closed system including the measuring cells and the preparational units. Details of the continuous deionization procedures have been given elsewhere [30, 31]. Since preparation may have severe influence on electrokinetic measurements, we nevertheless here give a short outline of the preparation set-up. The suspension is pumped peristaltically through a closed Teflon® tubing system connecting i) an ion exchange chamber; ii) a reservoir under inert gas atmosphere to add further suspension or salt solutions, if electrolyte concentration dependent measurements are performed; iii) the conductivity measurement; and iv) a cell for static light scattering or transmission experiments. *Via* the static structure factor or the turbidity the latter facilitates an *in situ* control of $\Phi$, respectively the particle number density $n = \Phi / (4\pi/3) a^3$ where a is the particle radius [32]. Control of n may also be performed via the conductivity at completely deionized conditions (see below). Relative uncertainties in the packing fraction $\Phi$ are typically below 2% at $\Phi = 10^{-3}$. We note, that if the precleaning is performed carefully no aggregation is induced by peristaltic pumping. The whole system (excluding the pump) may be thermostatted to ±0.2 °C. Conductivity is measured at a frequency of $\omega$ = 400Hz [electrodes LTA01 or LTA1 and bridge WTW 531, WTW, Germany]. To check for reproducibility we compared conductivity values at different frequencies $\omega \leq$ 1kHz but found no dependence on $\omega$. In general the reproducibility of measurements was found to be better than 2%. Residual uncertainties were mainly due to contamination at low packing fraction and low salt concentrations.

During the experiments the ion exchange chamber is bypassed and great care is taken to assure stable experimental conditions on a typical time scale of a few hours. Leakage of stray ions into the system was estimated from an increase of the conductivity of pure water (55 nS/cm) of less than 30 nS/cm per hour in the electrophoretic cell to correspond to an NaCl equivalent of $5 \times 10^{-7}$ mol l$^{-1}$ h$^{-1}$. Another source of ionic impurities generally is provided by the particles themselves. The corresponding rise in conductivity was found equivalent to the addition of up to $10^{-6}$ mol l$^{-1}$ h$^{-1}$ of NaCl. Due to these effects we estimate an upper bound for unidentified small ion concentrations during experiments on „completely deionized" samples to be between $1 \times 10^{-7}$ and $5 \times 10^{-7}$ mol l$^{-1}$.

Before filling the suspension into the tubing system, the latter is rinsed with doubly destilled, filtered water through both the ion exchange column and the bypass until the eluate shows a conductivity below 60nS/cm. The volume of water in the tubing system depends on the arrangement of components. It is on the order of 40 cm$^3$ and is exactly determined by weighing. Then an Ar-atmosphere is laid on top of the water surface in the reservoir and stock suspension is added to adjust the desired volume fraction. The suspension is pumped through the ion exchange column for some ten minutes during which conductivity reaches a constant low value. Complete deionization ($c < 10^{-7}$ mol l$^{-1}$) is reached after roughly one hour. If desired, then a certain amount of salt is added. A few minutes of further pumping are generally sufficient to reach a constant but higher conductivity value. We note that for each measurement the suspension is deionized again, before a higher amount of salt or further particles are added and the procedure is repeated. In each case, changes in the suspension volume are accounted for.

During conductometric titrations the contamination with neutral $CO_2$ significantly influences the conductivity as it is neutralized in parallel with the surface groups:

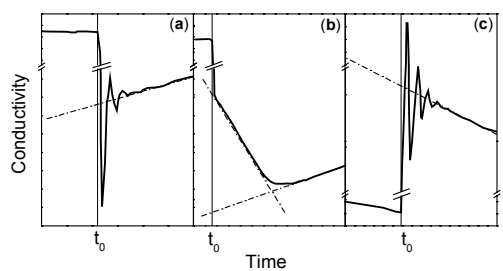

*Fig. 1*: Examples of conductivity/time traces taken (a) before. (b) slightly past and (c) far past the equivalence point. Due to contamination with neutral $CO_2$ the conductivity does not reach a constant value even after the decay of the oscillations due to mixing. $t_0$ defines the time of addition of NaOH.

usually no stationary value of $\sigma$ is observable. Examples of conductivity/time traces are given in Fig. 1 (a)-(c). The three data sets correspond to different amounts of added base: (a) is taken before, (b) slightly past and (c) far past the surface group equivalence point. In contrast to the case of acids, here, the conductivity does not reach a constant value even after the decay of the oscillations due to mixing. Instead σ rises (falls) steadily in samples before (far past) the equivalence point. Slightly past the equivalence point it first decreases to then increase again. We interpret this finding as indicating of an underlying two step chemical reaction in which first $CO_2$ reacts to give small amounts of free $H_2CO_3$ which is then neutralized to yield sodium carbonate. In this second step the particle acts as base and as a result is protonated again. Consequently the conductivity will change due to two contributions: the reversal of the neutralization reaction and the addition of sodium carbonate. If σ is recorded over some period of time after the addition and the value at t = 0 is extrapolated back, complete deionization after each addition is not necessary. Instead the change in conductivity is converted to added Carbonate and monitored throughout the experiment. The equivalence point is then inferred from the corrected conductivities as is shown in Fig. 2.

FEP(PFOA)78 and PS(SDS)102 also were investigated using this tubing system, again exploiting the possibility to perform several different *in situ* measurements simultaneously on one sample. Here, however, a very slow but irreversible desorption of surfactant takes place parallel to the deionization process. For FEP(PFOA)78 the number of adsorbed ionogenic groups was found to decrease by an order of magnitude within roughly four days of contact with IEX. Thus the desorption/discharging kinetics under neutral conditions are slow enough to consider the charge to be constant during the conductivity measurements. As a further precaution we did not deionize the sample after each addition of salt but corrected for contamination with stray ions via measurements of the temporal evolution in σ. We note, however, that upon addition of NaOH our measurements seem to indicate an enhanced desorption of surfactant from the surface which is subsequently removed upon new deionization (see also below).

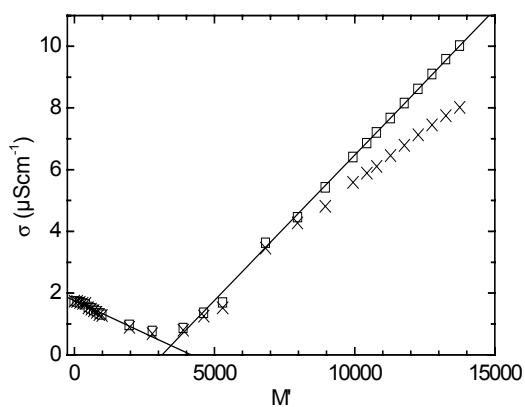

*Fig. 2*: Titration curve for PS115. Data are shown as a function of M´, the number of $Na^+$ added per particle. Crosses: raw data; open squares: data corrected for $CO_2$ contamination. The solid lines are linear fits to the corrected data.

Previous measurements on PS(SDS)102 with extensive IEX contact showed desorption of larger amounts of SDS leading to coagulation [31]. In the present study we therefore used PS(SDS)102 in the dialized state without further contact with IEX, i.e the IEX chamber was bypassed in all measurements.

To determine the electrophoretic mobilities μ of the particles a conventional Doppler velocimetry with real space moving fringe illumination and incoherent detection followed by FFT-frequency analysis [Ono-Sokki, NTD, Japan] was used. Details are given elsewhere [31]. The velocities v were found to increase linearly with the applied field strength E in all cases and no frequency dependence of mobilities μ = v / E was observed. Under our experimental conditions the residual uncertainties are less than 10%. Measured mobilities of particles under consideration are in the range of  μ = (1.5-8) $10^8$ $m^2$ $s^{-1}$ $V^{-1}$, show a pronounced Φ dependence at low packing fraction but saturate at larger values of Φ [31], for FEP(PFOA)78 see also [29].

## Results

Figs. 3 to 5 show the dependence of measured conductivities over a wide range of packing fractions for the different samples.

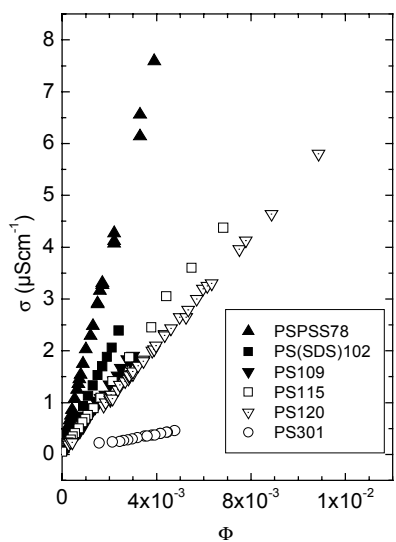

*Fig. 3*: Packing fraction dependence of conductivities for different deionized suspensions.

From Fig. 3, where data are plotted against the packing fraction it is evident that in all cases the conductivity depends linearly on $\Phi$ but with different proportionality factors. To check for the range of linearity we plot the conductivity of sample PS115 corrected for the theoretical conductivity of the background $\sigma_{H2O}$ in a double logarithmic way against the particle number density n in Fig. 4. The data arrange on a straight line of slope one down to n ≈ 0.1 $\mu m^{-3}$ ($\Phi$ ≈ 0.0001), where the particle contribution is on the order of the background contribution itself. The statistical accuracy at such low packing fractions is not sufficient to trace the validity of a linear packing fraction dependence to lower concentrations. The main source of error is obvious from the systematic deviation of $\sigma$ towards larger values. It indicates that in this range spurious amounts of stray ions (even of c = $10^{-7}$ mol $l^{-1}$) may severely alter the conductivity. In conclusion, however, linearity is confirmed as long as the particle contribution is larger than the background contribution. A second check is performed at higher packing fraction, where due to increasing pair interaction a phase transition from fluid to crystalline order, respectively between different crystal structures, occur.

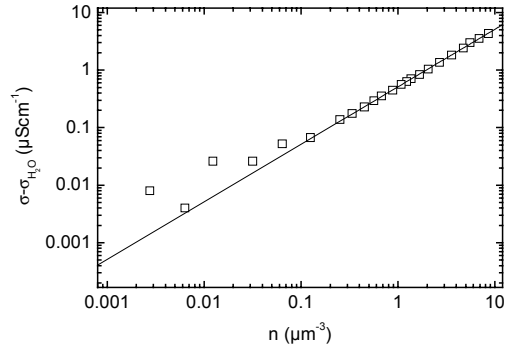

*Fig. 4*: Double logarithmic plot of the conductivity of PS115 corrected with the theoretical conductivity of the water $\sigma_{H2O}$ versus the particle number density n. The solid line is a fit of slope one to the data at n > 0.1 $\mu m^{-3}$. Deviations at lower n are due to contamination with spurious stray ions.

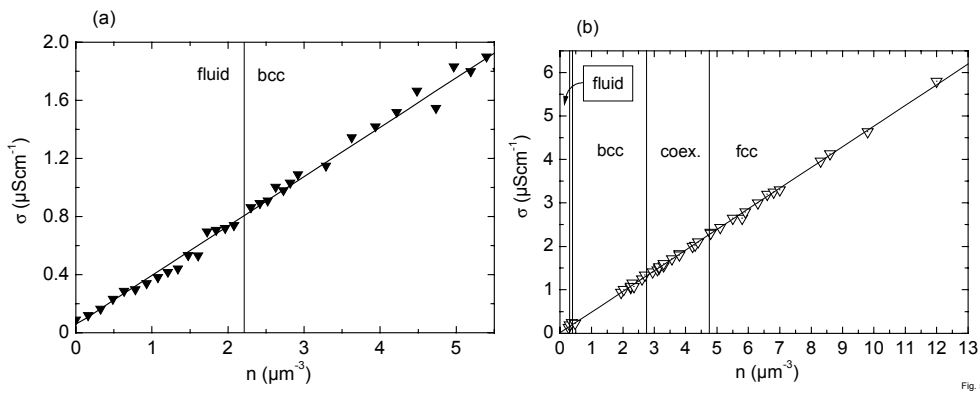

*Fig. 5*: Linearity of $\sigma(n)$ across the fluid-body centred cubic (bcc) and across the bcc-face centred cubic (fcc) transition for (a) PS109 and (b) PS115.

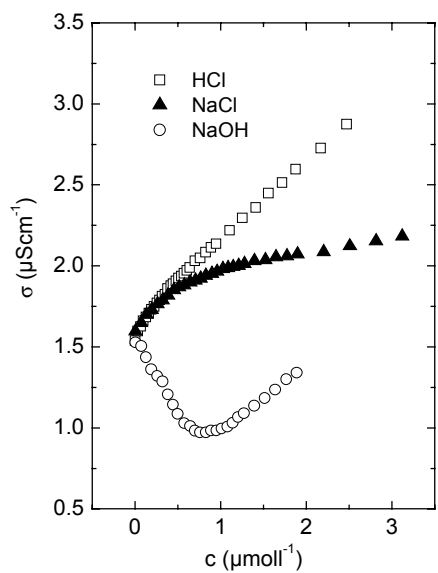

*Fig. 6*: Dependence of the conductivity on the kind of electrolyte used for PS109 at n = 4.36 μm$^{-3}$. Note that also upon addition of neutral electrolyte a pronounced non-linearity in σ is observed.

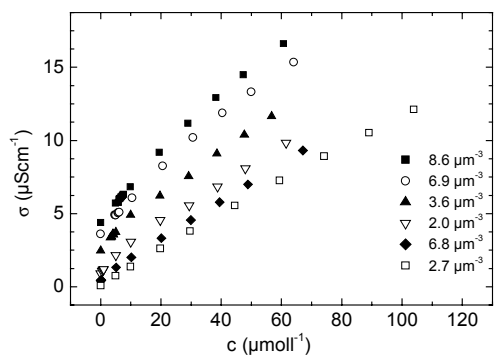

*Fig. 7*: Conductivity of PS115 as a function of NaCl concentration for different particle number densities n.

As presented in Fig. 5 (a) and (b) neither a change in the slope nor discontinuities across the transitions is observed within experimental error.

The same was observed also for other highly charged samples showing crystallization at low packing fraction and under deionized conditions.

In Fig. 6 we compare the influence of the addition of different electrolytes on the conductivity. We used a sample of PS102 at $\Phi = 0.00375$. $\sigma$ rises linearly for HCl. Adding NaCl $d\sigma/dc$ first is close to the value for HCl but then bends over to a lower slope. For selected measurement series on PS115 Fig. 7 demonstrates that this effect is present independent of packing fraction but the crossover shifts to lower values of c for decreasing $\Phi$. Also the appearance of the first steeper slope is less pronounced. Our findings confirm and systematically extend the results of Sumaru *et al.* and of Swetslot and Leyte to a much broader range of n and c. [15, 16]. They show that the non-linearities are not restricted to large packing fraction but a general feature of colloidal conductivity.

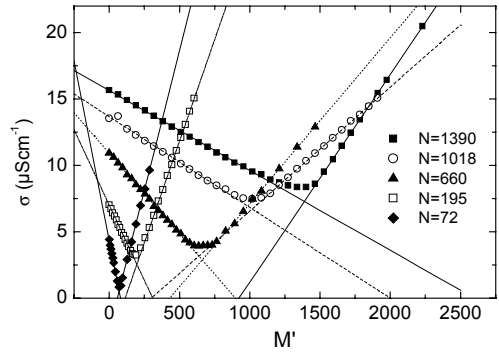

*Fig. 8*: Selected titration curves for FEP(PFOA)78. Data are shown as a function of M´. the number of Na$^+$ added per particle. With increasing contact time with the ion exchange resin the number of physisorbed surfactant molecules N decreases about one order of magnitude.

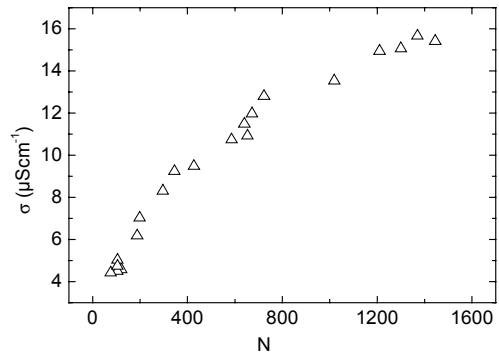

*Fig. 9*: Conductivity of deionized FEP(PFOA)78 samples at constant number density of n = $6.3 \times 10^{19}$ m$^{-3}$ as a function of the subsequently titrated surface group number N. A sublinear increase is observed.

For NaOH a titration curve typical for strong acids is observed [18, 19]. Data in Fig. 2 and 6 were taken successively and corrected for $CO_2$ contamination. For the PS samples investigated the corresponding numbers of surface groups N are compiled in Table 1. Using the charge variable FEP(PFOA)78 stabilized by physisorbed surfactant of low acidity we further checked on the systematic changes occurring for the titration curves in dependence on the surface group number. A selection of titration curves is shown in Fig. 8. Data were taken with complete deionization before each new addition of NaOH. After increasing time of contact with IEX we observe a pronounced change in the number of ionizable groups N as inferred from the point of equivalence. This decrease of N with decreasing conductivity has been presented already in [14]. In addition, we here observe the steepness of the initial slopes to increase as N gets smaller. It also increases as the packing fraction is increased (data not shown).

In Fig. 9 shows the conductivities of FEP(PFOA)78 under conditions of complete deionization as a function of the subsequently titrated group number N. Data are taken from four independent measurement series. Within each series the particles became less and less charged during subsequent titration deionization cycles. The series were aborted, when at very low charge coagulation occurred. In Fig. 9 the conductivity increases sublinearily. As will be shown below this can be understood as due to the combined effects of dissociation equilibrium and charge renormalization.

**Conductivity model**

A few years ago Deggelmann *et al*. proposed independent migration of all ionic species present in the solution [2111]. They used the empirical formula:

$$\sigma = neZ_\sigma^* (\mu_P + \mu_{H+}) + c\lambda_\infty + \sigma_B \tag{1}$$

where the molar conductivity of the added electrolyte at infinite dilution is given by $\lambda_\infty$ and the molar concentration by c. The background conductivity of unidentified small ions is denoted $\sigma_B$, e is the elementary charge, $\mu_P$ and $\mu_{H+}$ are the mobilities of particles (measured) and protonic counter ions, respectively. All deviations from ideal additivity are condensed into an effective charge $Z^*_\sigma$. In using an effective charge Eqn. (1) follows the pioneering work of Schaefer [10] but explicitly proposes a particle contribution to conductivity.

The salt concentration dependence, as given in Eq. (1) for infinite dilution as $d\sigma/dc = \lambda_\infty$, does not apply to our data. Figs. 6 and 7 give clear and systematic evidence, that a pronounced nonlinearity in $d\sigma/dc$ is present even at $c \leq 10^{-5}$ mol l$^{-1}$. Previous authors were able to describe similar data on the basis of a numerically complex theoretical approach. We here are explicitely interested in an equally well performing but far more simple description. We therefore shall continue to use Deggelmanns basic assumption of independent ion migration to derive an alternative expression.

We consider the EDL to be devided into an outer part containing $Z^*_\sigma$ counterions of bulk mobility $\mu_+$ and an inner part containing $Z-Z^*_\sigma$ immobilized counterions. Immobilization may be due to several effects: hairy surface layers [33], non-specific ion association [12, 19] or slowing of small ion diffusion in the vicinity of the particle surface due to some kind of increased small ion-small ion interaction at elevated $c(r)$ [15, 16, 19]. Counterion condensation as described by renormalization concepts [7, 8, 9] is based on a electrostatic „binding" of dissociated counter-ions. If the latter effect dominates, one expects the numerically calculated renormalized charges $Z^*_{DH}$ to be very close to the effectively transported charges in a conductivity experiment. Choice of such a description does not mean that the counterions in the inner shell are completely immobile, rather it states that for a certain number $Z-Z^*_\sigma$ the translational degrees of freedom are coupled to the particle. These counter-ions move with the particle, while $Z^*_\sigma$ move independently. We note that a similar description is proposed in the dynamic Stern layer model of electrophoresis, where polarization effects behind the (electrophoretic) plane of shear are explicitely accounted for [34].

We further allow for an exchange of ions between the inner and outer part of the EDL, as long as the overall radial charge distribution is retained. If now salt is added, salt ions may exchange with counterions of equal charge sign, while coions are assumed to stay outside the proposed inner shell due to electrostatic repulsion. It further is convenient to introduce the number concentration $M = c\ 1000\ N_A / n$ of small ions per particle and arithmetic mean small ion mobilities:

$$\bar{\mu}^+ = \frac{\mu_i M_i^+}{M_i^+} \quad ; \quad \bar{\mu}^- = \frac{\mu_i M_i^-}{M_i^-}. \tag{2}$$

We note that the sum is over all small ions actually present i.e. including the Z counter-ions. We further note that in the case of chemical reactions the number of excess small ions *present* per particle M does not necessarily equal the number M´ *added* per particle. Assuming as before additivity of all conductivity contributions ($\sigma = \Sigma\, n_i\, e\, z_i\, \mu_i$ with $z_i = 1$ in our case of monovalent salt and counter-ions) we may formulate:

$$\sigma = ne\left(Z_\sigma^*\left(\mu_P + \bar{\mu}^+\right) + M\left(\bar{\mu}^+ + \bar{\mu}^-\right)\right) + \sigma_B \qquad (3)$$

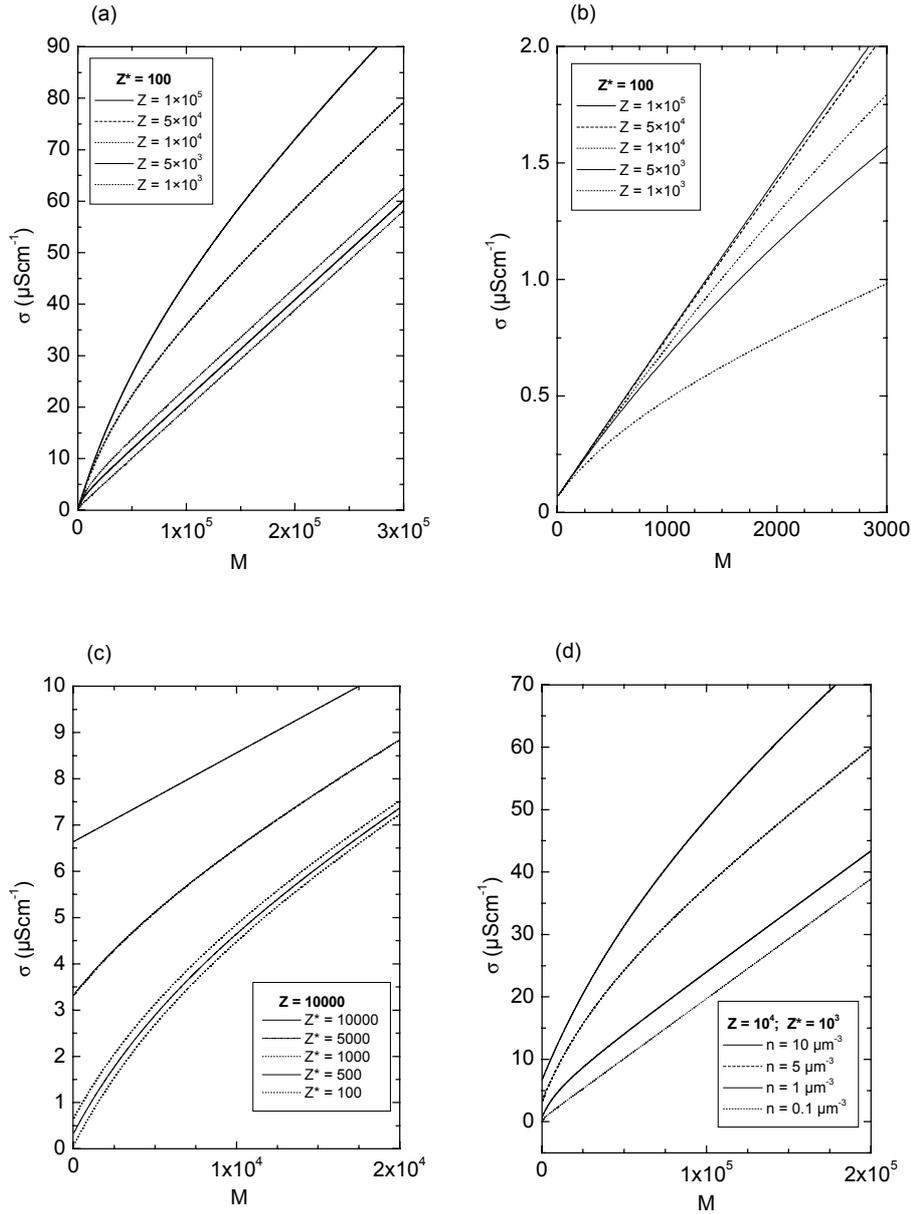

*Fig. 10*: Results of numerical calculations using Eq. (2) and (3) for the conductivity as a function of neutral electrolyte concentration expressed in terms of M, the number of Na$^+$ *present* per particle. Calculations (a)-(c) at fixed particle number density n = 1 μm$^{-3}$. (a) Fixed effective charge Z*. bare charge Z decreasing from top to bottom; (b) the same but shown over a larger range of M; (c) Fixed bare charge Z. effective charge decreasing from top to bottom; (d) both charges fixed and n decreasing from top to bottom.

While this concept is actually straight foreward and simple, the resulting combined dependency on M and on n is quite complex. To illustrate this we show a number of numerical calculations in Fig. 10 (a) to (d). First the surface charge Z is varied at constant Z* and fixed n. Fig. 10 (a) shows the dependence of conductivity over a large range of M, Fig. 10 (b) shows the range up to M = 3000 enlarged. From Fig. 10 (b) it would seem that the uppermost curve is linear, while in Fig. 10 (a) the lowest curve has a linear appearence. However, all curves start with a large slope and show a crossover to a lower slope. With decreasing charge ratio Z*/Z the crossover region shifts to larger values of M.

A strictly linear behaviour with a slope corresponding to the addition of NaCl to pure water is encounterd only, if the charge ratio is 1, as is the case for the uppermost curve in Fig. 10 (c). With decreasing charge ratio the initial slope approaches that expected for the addition of HCl to pure water. Finally in Fig. 10 (d) the dependence on particle density n is shown at fixed charge and charge ratio. The resulting curves start at increasing conductivity values; all are bent with the crossover region shifting to larger M for increasing n.

**Discussion of low frequency conductivities**

*Packing fraction dependence.* In the case of deionized suspensions Eqn. (3) reduces to:

$$\sigma = \sigma_0 + \sigma_{H2O} + \sigma_B = neZ^*_\sigma(\mu_P + \mu_{H+}) + \sigma_{H2O} + \sigma_B \tag{4}$$

As can be seen from Fig. 3 the predicted linear dependence on n is fulfilled for all curves. $Z^*_\sigma$ is taken as only fit parameter and our results are compiled in Table 1. In fitting Eqn. (4) in Figs. 4 and 5 we make use of the fact that for sufficiently large n the measured particle mobilities become independent of n at values around (5-10) $\mu ms^{-1}/Vcm^{-1}$ [31]. For comparison the mobilities of $H^+$, $OH^-$, $Na^+$ and $Cl^-$ are $36.5 \times 10^{-8}$ m²/Vs, $15.8 \times 10^{-8}$ m²/Vs, $5.02 \times 10^{-8}$ m²/Vs, and $7.1 \times 10^{-8}$ m²/Vs, respectively [35]. The particle contribution therefore should not be neglected. $\sigma_{H2O}$ is calculated self consistently via the dissociation product of water with $-\log_{10}(c_{H+} \, c_{OH^-}) = 14$ and $c_{H+} = n \, Z^*_\sigma / 1000 \, N_A$. This contribution in principle becomes important for very low packing fractions, but in practice it usually is masked by $\sigma_B$.

*Variation of surface charge.* We recall that reduced effective values for electrokinetic charges were frequently employed also by other authors: Okubo e.g. suggested a description in terms of an effective dissociation coefficient (corresponding to a drastic shift in surface pH [36, 37]). Alternatively one may formulate the decrease in charge number in terms of an association equilibrium, where counter ions bind loosely to the particle surface ($pK_{NaAss} \approx 4.5$) and reduce the surface charge to a value $Z^*_{Ass} < Z$ [12, 19]. Only few studies are available with systematic variation of the particle charge under conditions of fixed particle density and particle size [29, 38]. In a recent paper Yamanaka *et al*. found the effective charge density as derived from conductivity to slowly but continously increase [39]. In this case the charge of silica particles was increased by addition of small amounts of NaOH, with the pH staying in the range of 6 to 8.

Also for our FEP(PFOA)78 sample it is seen from Fig. 9 that $\sigma_0$ keeps increasing with increasing N. We apply Eqn. (4) to these data to derive the effectively transported charge $Z^*_\sigma$ as a function of the titrated group number N. In Fig. 11 we compare the results to the case of $Z^*_\sigma = Z$. The continuous increase in the effective charge can be monitored up to values of $Z^*_\sigma = 370$. The slope decreases but no strict saturation is observed. As will be discussed below this can be traced back to the combined action of surface dissociation equilibrium and charge renormalization.

*Addition of neutral electrolyte.* Introducing the concentration of small ions per particle M allows to treat both counter-ions and added electrolyte on the same footing. Eq. (3) predicts a linear dependence on n irrespective of the electrolyte composition. Thus we plot in Fig. 12 the conductivity per particle versus M. All measurement series covering more than five orders of magnitude in M and more than three orders of magnitude in n fall on a single master curve. The low M limit of that curve corresponds to the conductivity contribution of a single particle and the surrounding EDL. Consistent with the observations in Figs. 3 to 5 this value does not change as a function of n.

To isolate the contribution of added electrolyte we rewrite Eq. (3) to read

$$\sigma = \sigma_0 + ne\left(Z^*_\sigma\left(\bar{\mu}^+ - \mu_{H+}\right) + M\left(\mu_{Cl-} + \bar{\mu}^+\right)\right) + \sigma_B \tag{5}$$

with the particle contribution $\sigma_0$ defined in Eq. (4). Neglecting $\sigma_B$ the first part of the electrolyte contribution goes to zero, whenever $\bar{\mu}^+$ equals $\mu_{H+}$.

From our model calculations this will be the case at small charge ratio $Z^*/Z$ and in the limit of $M \ll Z$. Then the initial slope in a plot of $(\sigma-\sigma_0)/n$ versus $M$ will correspond to that expected for the addition of HCl to pure water. Upon further addition of neutral electrolyte the number of counter-ions is conserved and only their visibility changes as expressed through Eq. (2). We show this in Figs. 13 to 15. In Fig. 13 $(\sigma-\sigma_0)/n$ is plotted versus M for large and small M (insert). For large M the data follow the conductivity increase expected for the addition of NaCl to pure water, while at low M the conductivity values are smaller and a larger slope is observed.

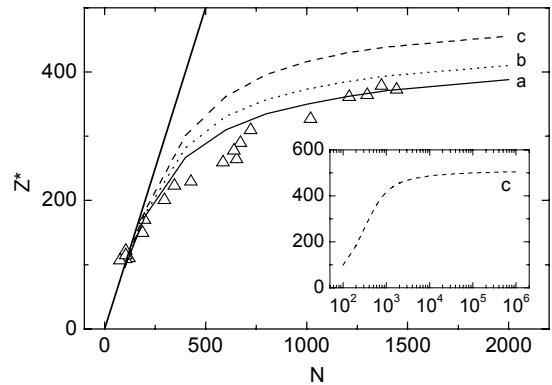

*Fig. 11*: Effective charges as a function of titrated surface group number for FEP(PFOA)78. Open symbols: $Z^*_\sigma$ as derived from conductivity under deionized conditions; bold straight line: $Z^* = N$ (linear increase); Calculations performed for $\Phi = 0.016$; $a = 39$nm and a) pK = 4.0. $c_B = 2\times10^{-7}$ mol$^{-1}$l$^{-1}$; b) pK = 4.0. $c_B = 1\times10^{-6}$ mol$^{-1}$l$^{-1}$. and c) pK = 2.3. $c_B = 2\times10^{-7}$ mol$^{-1}$l$^{-1}$. The insert shows $Z^*$ calculated for case c) over a larger range of ionizable groups N. Saturation is observed for $N > 10^5$.

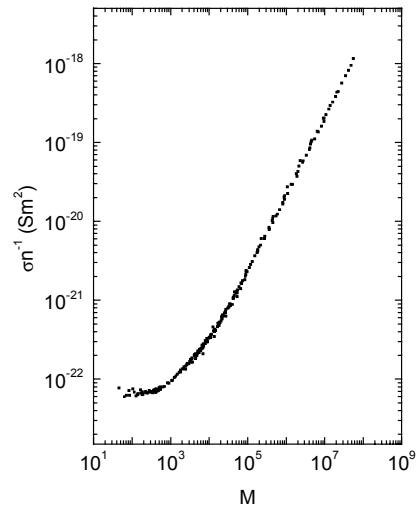

*Fig. 12*: Conductivity contribution per particle as a function of M. For all measurements performed on PS115 over a wide range of c and n the data fall on a single curve.

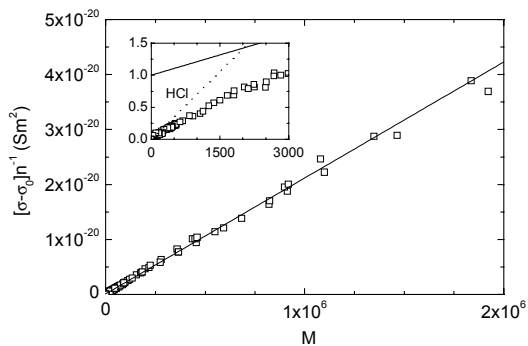

*Fig. 13*: Conductivity contribution of the added electrolyte for large M. The solid line is a fit of Eq. (1) for NaCl to data at M>100. The insert shows significant deviations to occur at small M leading to an unphysically high background contribution.. Also shown is the expectation for HCl according to Eq. (1) with $\sigma_B = 0$.

The double logarithmic plot in Fig. 14 shows the gradual transition between the two limiting linear behaviors and compares them to the model calculations. The model is able to qualitatively describe the data over the whole range of measurements. To obtain a quantitative description we fix the effective charge to the value obtained from the n dependent measurements and leave Z as the only free parameter. The best fit to all data in Fig. 14 yields Z = 2830. As can be seen, however the experimental data are somewhat below the fit curve at low M. In Fig. 15 we use a linear scale and compare to three model calculations. The solid line is the same as in Fig. 14, the other two lines correspond to a 25% larger (smaller) charge. The lower charge fits the data at low M better than the original fit curve. This possibly reflects an increased dissociation due to the shift of surface pH as the counter-ions are exchanged and the composition of the electrolyte is changed or a change in the effective charge as a function of M. Both is not yet implemented in our model which assumes constant Z and $Z^*_\sigma$. On the other hand our investigation clearly shows that both charge numbers are relevant for a quantitative description of conductivity.

*Titration with NaOH*. Upon addition of NaOH to a deionized suspension with low surface pK the protonic counter-ions are neutralized by the hydroxyl-ions and replaced by sodium counter-ions. Thus for low amounts of base added, the number of counter-ions, respectively the number of dissociated surface groups does not change. As long as there is no excess electrolyte, M stays zero. Beyond the equivalence point $M = M\`- N$, where $M´$ is the number of small ions added. $M´$ also enters the calculation of mean ion mobilities. For the initial decrease Eq. (3) now reads:

$$\sigma = neZ^*_\sigma(\mu_P + \overline{\mu}^+) + \sigma_B \tag{6a}$$

with:

$$\bar{\mu}^+ = \frac{(Z-M')\mu_{H+} + M'\mu_{Na+}}{Z} = \mu_{H+} + \frac{M'\mu_{Na+}}{Z} - \frac{M'\mu_{H+}}{Z} = \mu_{H+} + \frac{M'}{Z}\Delta\mu \quad (6b)$$

where the mobility difference $\Delta\mu = \mu_{Na+} - \mu_{H+}$ has been introduced. Insertion in Eq. (6a) yields:

$$\sigma = \sigma_0 + neM'\frac{Z^*_\sigma}{Z}(\Delta\mu) + \sigma_B \quad (7)$$

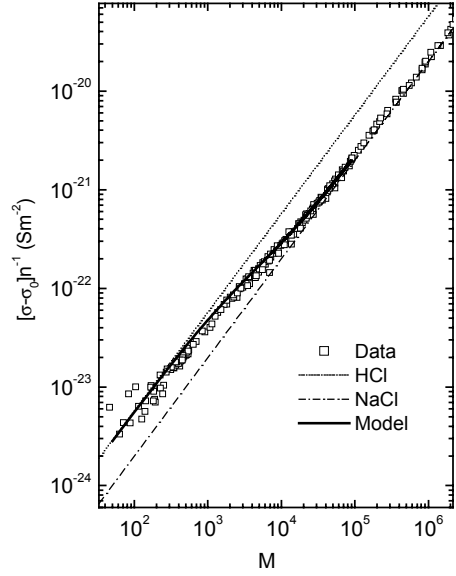

*Fig. 14*: The same data plotted double logarithmically demonstrating the transition between the two limiting linear regimes. The bold line is a fit of our model to data points past the transition.

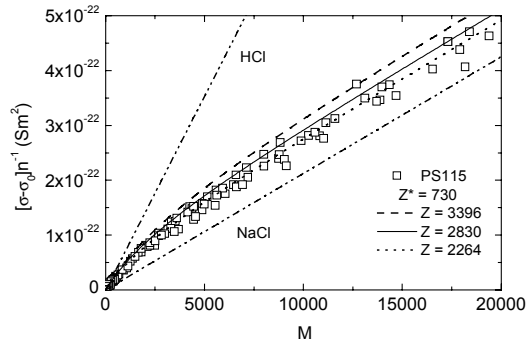

*Fig. 15*: Model calculations using Eq.(2) and (3) for $Z^*_\sigma = 730$. The solid line corresponds to the fit in Fig. 14. At low M our data are better described. if a slightly smaller bare charge is used. Also shown are the expectations for HCl (uppermost curve) and NaCl (lowest curve) according to Eq. (1) with $\sigma_B = 0$.

The initial slope $d\sigma/dM' = en\Delta\mu Z^*_\sigma/Z$ depends not only on the small ion mobility difference and the particle number density n but, most importantly, also on the charge ratio $Z^*_\sigma/Z$. As $Z^*_\sigma$ is already known from $\sigma_0$ we may extract Z from the limiting slope. A single titration curve thus may yield all three relevant charge numbers N, Z and $Z^*_\sigma$. The same holds at larger surface pK for very small M', i.e. the first steeper decrease observed in the titration curves of week acids. Then, however, the excess base needed to cause full dissociation has to be accounted for, if calculating the conductivity at larger M'.

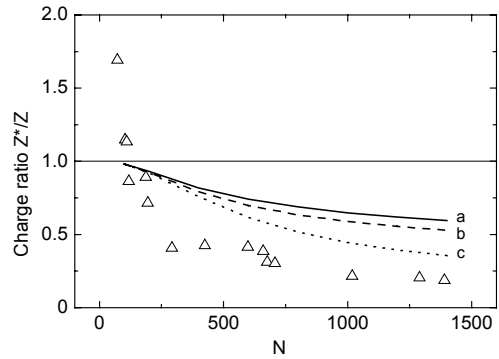

*Fig. 16*: Comparison of charge ratios Z*/Z as measured for FEP(PFOA)78 (open symbols) to calculations for $\Phi = 0.016$; a = 39nm and a) pK = 4.0. $c_B = 2\times10^{-7}$ mol$^{-1}$l$^{-1}$; b) pK = 4.0. $c_B = 1\times10^{-6}$ mol$^{-1}$l$^{-1}$. and c) pK = 2.3. $c_B = 2\times10^{-7}$ mol$^{-1}$l$^{-1}$. The charge ratio is observed to drop with increasing background concentration and decreasing pH. As compared to Fig. 11 here curve a gives the worst data description. For discussion see text.

Fig. 16 shows the charge ratio obtained from the initial slopes of the FEP(PFOA)78 titration curves plotted as a function of N. The charge ratio drops to low values reaching approximately 0.2 at N = 1400 thus confirming the trend already visible from Fig. 11.

Unexpectedly, however, the first two data ponts lie above unity. This cannot be explained by experimental uncertainties in determining the slope (c.f. Fig. 8). A possible explanation may be the assumption of desorption of ionic groups from the surface upon exchange of the counter-ion. The perfluorinated acid indeed is not water soluble in the protonic form, while it dissolves to a small amount, if in the sodium form. Free surfactant would not shift the equivalence point, since no deionization is performed during the titration. Free surfactant, however will have a mobility $\mu_S$ different to that of the particle. In the case of $\mu_S > \mu_P$ the contribution from the surfactant may overcompensate the reduction in charge and mobility of the particle. At larger particle charges this contribution would not be visible due to the restricted solubility of the surfactant, even in the sodium form.

Far past the equivalence point all protons are neutralized and:

$$\sigma = neZ_\sigma^*(\mu_P + \mu_{Na+}) + M(\mu_{OH-} + \mu_{Na+}) + \sigma_B \qquad (8)$$

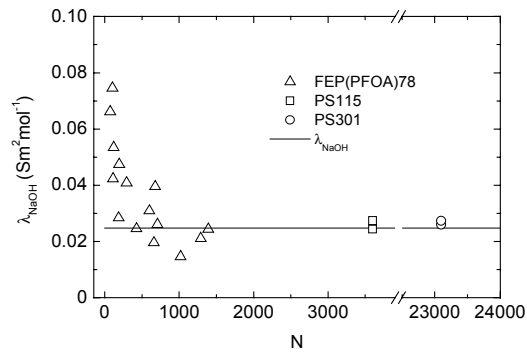

*Fig. 17*: Limiting molar conductivity of NaOH as derived from the slope of titration curves past the equivalence point. Open Squares: EP(PFOA)78; closed circles: PS115; Closed squares: PS301; solid line reference value for aqueous NaOH at infinite dilution and T = 25°C. While the reference value is obtained for all PS samples it is approached by FEP(PFOA)78 only for N > 500.

No nonlinearity results, as counter-ions and further added cations are of the same kind. In fact, the slope should correspond to the addition of NaOH to pure water, irrespective of the particle charge. In Fig. 8, however a clear dependence of the final slopes is visible. In Fig. 17 we plot the corresponding molar conductivities for the FEP(PFOA)78 and two PS samples and compare them to the literature value of $\lambda_{\infty,NaOH} = 247.7 \times 10^{-4}$ S m$^2$ mol$^{-1}$ [35]. For FEP(PFOA)78 the data approach the expected value only for large N, while in the case of PS even at a relatively large charge ratio the literature value is met within the experimental errors. Also here the deviation at small N cannot be explained within our model without any further asumption. Allowing, however, for desorption of S surfactant molecules per particle at sufficiently large pH and further assuming the charge ratio at low N to be close to one, we rewrite Eq. (8) to get:

$$\begin{aligned}\sigma &= ne[(Z-S)(\mu_P + \mu_{Na+}) + S(\mu_S + \mu_{Na+}) + M(\mu_{OH-} + \mu_{Na+})] + \sigma_B \\ &= ne[Z(\mu_P + \mu_{Na+}) + S\Delta\mu + M(\mu_{OH-} + \mu_{Na+})] + \sigma_B\end{aligned} \quad (8a)$$

Where $\Delta\mu = \mu_S - \mu_P$ is the mobility difference between particle and surfactant. Thus the final slope will increase above the values expected for the addition of NaOH to pure water proportional to the amount of desorbed surfactant. Given the limited solubility of the surfactant even in the sodium form, it is clear that at large Z the ratio S/Z stays small and the correction becomes negligable. If this explanation holds further tests with surfactants of known mobility, this would be an elegant way of directly studying the adsorption/desorption kinetics of colloid/surfactant systems. In addition the presented interpretation would support the importance of including both small and macro-ions in the description of conductivities.

*Comparison of effective charges.* There exist several approaches to numerically calculate bare charge numbers Z from titrated group numbers N, to calculate the structure of the double layer under boundary conditions of constant surface potential or charge and to calculate effective charges Z*. Notice, that all effective charge numbers used in this paper are indicated by a star *. In particular the electrostatic potential φ(r) and the structure of the EDL forming around a colloidal particle have comprehensively been studied by theoretical and numerical work. It turned out that for most practical purposes a mean-field description by solutions of the non linearized PB solved numerically within a spherical Wigner-Seitz cell (PBC-model [7]) yields consistent results to more sophisticated approaches [8, 9]. For the prediction of interaction dependent properties, like phase behavior or elastic moduli a fit of a Debye-Hückel (DH) potential to φ(r) at the cell boundary is performed with a renormalized charge and a renormalized screening constant Z*$_{DH}$ and κ*$_{DH}$ as free parameters. Again consistent results are obtained for different prescriptions and techniques in the limit of large particle radii as compared to the Bjerrum length $\lambda_B = e^2 / 4\pi \varepsilon k_B T$ which in water is 0.72 nm [7, 8, 9]. Here $\varepsilon = \varepsilon_0 \varepsilon_r$ the dielectric permittivity of the suspending medium and $k_B T$ the thermal energy. In many cases the effective charges $Z^*_{DH}$ derived this way were found to be numerically very close to those derived from interaction dependent equilibrium properties. Thus the charge renormalization concept is widely accepted in the description of phase behaviour, structure formation, diffusivity or elasticity data. In the case of electrophoretic transport it has recently been shown that there are large quantitative discrepancies between calculated and measured effective electrophoretic charges [31, 36]. This raises the interesting question how $Z^*_{DH}$ compares to effective charges derived from conductivity [40].

We therefore calculated both the bare particle charge $Z_{PBC}$ and the renormalized effective charge $Z^*_{DH}$ using a program kindly provided by Luc Belloni, which is based on the PBC model under the boundary condition of constant surface dissociation equilibrium [41]. Charge numbers are calculated for the PS samples using a surface pK of 0.5, the titrated group number N and the radii as compiled in Table 1. For deionized samples the unidentified small ion concentration in the suspension was fixed to $2 \times 10^{-7}$ mol l$^{-1}$. Effective charges calculated this way were observed to be significantly smaller than the bare particle charges. Except for PSPSS70 the results compiled in Table 1 are quite close to the values of $Z^*_\sigma$ derived from fits of Eqn. (4) to conductivity data.

For FEP(PFOA)78 we performed calculations of Z and $Z^*_{DH}$ using the PBC-program with a radius of a = 39 nm, fixed particle number density of n = 63 μm$^{-3}$. Three calculations are shown in Fig. 11. In all cases the renormalized charges calculated steadily increase over the range of experimental N. We obtain a reasonably good agreement between calculated $Z^*_{DH}$ and measured $Z^*_\sigma$ (open squares) for pK = 4.0 and $c_B = 2 \times 10^{-7}$ mol l$^{-1}$ as estimated from the contamination experiments on pure H$_2$O. We note that such a large pK actually seems to contradict the form of our titration curves. If we use pK = 2.3 we obtain $Z^*_{PBC} \approx N$ and the upper curve for $Z^*_{DH}$.

We have to stress that the calculation is very sensitive to the choice of boundary conditions. In fact the curve a of Fig. 11 is practically identical to Fig. 4 in [14] calculated using a slightly different radius pK and background concentration. In Tab. 2 we compile some representative results to demonstrate the influence of variations in individual input parameters for the numerical calculations. In particular $c_B$ and the pK have a strong influence on the numericaly calculated $Z_{PBC}$. For fixed $Z_{PBC}$ the renormalized charge increases with increasing particle density but even more so with increasing radius. An increase in the background salt concentration is of little influence. The quantitative derivation of particle parameters from a fit of calculated renormalized charges to effectively transported charges thus has to be rated questionable for the present study. It rather remains an important challenge.

Qualitatively however, it is exactly the fact of incomplete dissociation, that produces the particular non-saturating shape of the experimental curve in Fig. 9. For sufficiently large number of dissociated groups Z, the predicted saturation of effective charges is observed in the numerical calculations. This is shown in the insert of Fig. 11 for curve c. Here saturation is reached around $N \approx 10^5$ where Z is of the order of a few times $10^3$. We note that similar Z values necessary for saturation of Z* are observed for the other two cases albeit at still larger N. Such a large Z was not reached in the experiment. We further note that in all three calculations for $n = 6.3 \ 10^{19}$ m$^{-3}$ the saturation value of $Z^*_{MAX} = 510 \pm 30 = 9 \ a/\lambda_B$ is in good agreement with the value proposed by Stevens [8] for vanishing packing fraction.

A comparison of charge ratios $Z^*_\sigma/Z$ as derived from the conductivity titration to the results from the numerical calculations is performed in Fig. 16. We show the data for $c_B = 2 \times 10^{-7}$ mol l$^{-1}$, a = 39nm, $\Phi = 0.0157$ and pK = 4.0. The charge ratio drops from one to values of about 0.2 around N = 1400 in good agreement with the experimental data at large N.

From these results we conclude that most of our conductivity data are not inconsistent with the charge renormalization concept and give at least some qualitative support. During our studies we learned, however, that our present results do not allow for a quantitative and decisive test as yet.

The major uncertainty in using the PBC program is the lack of knowledge on the exact pK of the acid and the background concentration $c_B$. The major uncertainty in using the conductivity model is the lack of sufficient data on the particle mobility upon addition of NaOH. In the present investigation the pK was determined from a fitting procedure, and $c_B$ was estimated, while the mobilities were assumed to retain their values at deionized conditions. In the light of the strong sensitivity of the numerical calculations it is not too surprising that also other authors have reported interesting deviations from specific predictions of the charge renormalization model. We recall that at very low $\Phi$ the experimentally determined effective electrophoretic charge was found to be much smaller than expected [31, 36, 37] and that a continued increase in $Z^*_\sigma$ with increasing N has been reported [39] where actually a saturation should occur. On the other hand saturation was unequivocaly observed for the effective charge as determined from measurements of the static structure factor of micellar systems [38]. It thus is necessary to extend the measurements to much larger N under precise control of the background concentration $c_B$ and at the same time vary the packing fraction for particles of well characterized surface chemistry, i.e. independently determined surface pK.

**Conclusions**

We have performed comprehensive investigations on the low frequency conductivity in mixtures of different simple 1:1 electrolytes and highly assymetric Z:1 electrolyte, where Z was on the order of $10^2$ to $10^4$. Electrolyte concentrations and particle number densities were varied over a wide range covering isolated particles as well as fluid or crystalline ordered suspensions.

Practically all our data could be described within a simple yet powerful conductivity model which assumes additivity of contributions of all species present. All macro ion small ion interactions are condensed into an effectively transported charge $Z^*_\sigma$ which was found to be rather close to the renormalized charged $Z^*_{DH}$ as obtained as fit parameter from the results of PBC-model calculations. This quantity is particle specific and independent of c and n.

The conductivity model assumes a division of the EDL into an inner non-conducting part and an outer part with bulk conductivities. However, to describe the data properly we extended the model already proposed by Deggelmann *et al.* to allow for an exchange of cationic species between the two parts. The model correctly predicts the packing fraction dependence. It correctly predicts the linearity in M for cases of no compositional change within the EDL, i.e. the addition of HCl and of NaOH past the equivalence point in titrations. Most importantly it also describes the pronounced non-linearities in M observed upon the addition of neutral electrolyte, i.e. NaCl.

From the titration curve for a macro ion of chemically bound charge the model is able to extract three relevant charge parameters under low salt conditions: the intersection of initial and final slopes yields the number of surface groups N; the conductivity in the deionized state yields the effectively transported charge $Z^*_\sigma$ and the initial slope yields the charge ratio $Z^*_\sigma/Z$ and thus the number of dissociated charges in the deionized state. For our titration data on FEP(PFOA)78 we observed charge ratios above 1. This led to an extension of the model under the assumption of partial desorption of surfactant which was supported also by the deviation of the slopes past the equivalence point towards larger than expected values. We presently conduct further experiments to explore this point in more detail. If the validity of Eq. (8a) could be confirmed in further studies this would in turn provide an elegant means of determining adsorption/desorption proceses of charged surfactants.

Combining these individual observations the most important result of our studies can be phrased as follows. The simple model proposed indicates that a description in terms of simple additivity is sufficient to capture the basic, but nevertheless complex features of conductivity in highly asymmetric electrolyte mixtures. The only free parameter of this description, the effectively transported charge is close to numerically calculated values. This opens the perspective of applying instrumentally simple conductivity measurements to the prediction of other suspension properties, once the charge renormalization concept has been further tested.

**Acknowledgements**

It is a pleasure to thank Luc Belloni, Roberto Piazza, Tommaso Bellini, Cecco Mantegazza, and Hartmut Löwen for many fruitful discussions. We are further indepted to the group of Vittorio Degiorgio and Roberto Piazza for the kind gift of FEP(PFOA)78 particles, the group of Matthias Ballauf for the kind gift of PS(SDS)102 and PSPSS70 samples and Luc Belloni for the PBC program, respectively. These collaborations were supported by the DAAD Vigoni program and the HCM network CHRX-CT9. Further, we gratefully acknowledge the financial support of the Bundesministerium für Bildung, Wissenschaft, Forschung und Technologie (BMBF) and the Materialwissenschaftliches Forschungszentrum (MWFZ) Mainz.

**Additional Tables**

Table 2: Results of charge calculations for FEP. Left side: input parameter; right side: results. Uppermost row: set of typical parameters. Middle part: calculation of the concentration of ions at the Wigner-Seitz cell boundary. the number of dissociated surface groups $Z_{PBC}$ and the renormalized charge $Z^*_{DH}$. We study the influence of increasing $c_B$. of increasing N. of increasing N and $c_B$. of decreasing the pK. of increasing the particle number concentration n and of increasing a. respectively. Varied parameters in italics. The results for the particle charge numbers depend strongly on the interplay between the different parameters. Lower part: calculation of $Z^*_{DH}$ for fixed $Z_{PBC} = 800$ and varied $c_B$. n and a (N and pK do not enter this part of the calculations). Here the results are very sensitive on the competition between radius and packing fraction; the salt concentration only marginally shifts $Z^*_{DH}$ to higher values.

| N | a / nm | Φ | pK | $c_B$ / μmol l$^{-1}$ | $c_{WS}$ / μmol l$^{-1}$ | $Z_{PBC}$ | $Z^*_{DH}$ |
|---|---|---|---|---|---|---|---|
| 800 | 39 | 0.016 | 4.0 | 0.2 | 28 | **487** | **335** |
| 800 | 39 | 0.016 | 4.0 | *1.0* | 30 | **563** | **357** |
| *2000* | 39 | 0.016 | 4.0 | 0.2 | 32 | **718** | **388** |
| *2000* | 39 | 0.016 | 4.0 | *1.0* | 33 | **871** | **410** |
| 800 | 39 | 0.016 | *2.3* | 0.2 | 32 | **768** | **396** |
| 800 | 39 | *0.024* | 4.0 | 0.2 | 42 | **499** | **341** |
| 800 | *45* | *0.024* | 4.0 | 0.2 | 31 | **548** | **384** |
| --- | 39 | 0.016 | --- | 0.2 | --- | 800 | **401** |
| --- | 39 | 0.016 | --- | *2.0* | --- | 800 | **402** |
| --- | 39 | *0.024* | --- | 0.2 | --- | 800 | **409** |
| --- | *45* | *0.024* | --- | 0.2 | --- | 800 | **451** |

# References

[1]  J. Lyklema: „*Fundamentals of Interface and Colloid Science, Vol. 1 & 2*", London, Academic, 1993

[2]  P. N. Pusey in J. P. Hansen, D. Levesque, J. Zinn-Justin: „*Liquids, freezing and glass transition*", 51st summer school in theoretical physics, Les Houches (F) 1989, Elsevier, Amsterdam 1991, pp. 763.

[3]  A. K. Sood: *Solid State Physics* **45**, 1 (1991).

[4]  M. O. Robbins, K. Kremer, G. S. Grest: *J. Chem. Phys.* **88**, 3286 (1988).

[5]  G. Nägele, *Phys. Reports* **272**, 217 (1996).

[6]  T. Palberg: *Curr. Opn. Colloid Interface Sci.* **2**; 607 -614 (1997).

[7]  S. Alexander, P. M. Chaikin, P. Grant, G. J. Morales, P. Pincus, D. Hone:*J. Chem. Phys.* **80**, 5776 (1984).

[8]  M. J. Stevens, M. L. Falk, and M. O. Robbins, *J. Chem. Phys.* **104**, 5209 (1996).

[9]  R. D. Groot, *J. Chem. Phys.* **94**, 5083 (1991).

[10]  D. W. Schaefer: *J. Chem. Phys.* **66**, 3980 (1977).

[11]  T. Palberg, W. Härtl, M. Deggelmann, E. Simnacher, R. Weber: *Progr. Colloid Polym. Sci.* **84**, 352 (1991).

[12]  L. P. Voegtli and C. F. Zukoski, *J. Colloid Interface Sci.* **141**, 92 (1991).

[13]  J. Yamanaka, H. Yoshida, T. Koga, N. Ise, T. Hashimoto: *Phys. Rev. Lett.* **80**, 5806 (1998).

[14]  T. Palberg, W. Mönch, F. Bitzer, L. Bellini, R. Piazza: *Phys. Rev. Lett.* **74**, 4555 (1995).

[15]  K. Sumaru, H. Yamaoka, K. Ito: *Ber. Bunsenges. Phys. Chem.* **100**, 1176 (1996).

[16]  J. P. H. Zwetsloot and J. C. Leyte, *J. Colloid Interface Sci.* **163**, 362 (1994).

[17]  J. Stone-Masui, A. Watillon: *J. Colloid Interface Sci.* **52**, 479 (1975).

[18]  M. A. Labib, A. A. Robertson: *J. Colloid Interface Sci.* **77**, 151 (1980)

[19]  R. O. James, J. A. Davis, J. O. Leckie: *J. Colloid Interface Sci.* **65**, 331 (1978).

[20]  D. A. Saville, *J. Coll. Interface Sci.* **71**, 477 (1979).

[21]  M. Deggelmann, T. Palberg, M. Hagenbüchle, E. E. Maier, R. Krause, Ch. Graf, R. Weber:
*J. Coll. Interface Sci.* **143**, 318 (1991).

[22]  T. Palberg, J. Kottal, F. Bitzer, R. Simon, M. Würth, P. Leiderer: *J. Colloid Interface Sci.* **168**, 85 (1995)

[23]  J. A. Davis, R. O. James, J. O. Leckie: *J. Colloid Interface Sci.* **63**, 480 (1978).

[24]  R. A. Robinson, R. H. Stokes: „*Electrolyte Solutions*", London, Butterworth, 1970.

[25]   H. Oshima, T. Healy, R. White: *J. Chem. Faraday Trans. 2* **77**, 2007 (1981).

[26]   B. Jönsson, H. Wennerström, P. G. Nilsson, P. Linse: *Colloid Polym. Sci.* **264**, 77 (1986).

[27]   V. Degiorgio, R. Piazza, T. Bellini, M. Visca: *Adv. Colloid Interface Sci.* **48**, 61 (1994).

[28]   U. Apfel, K. D. Hörner, and M. Ballauff, *Langmuir* **11**, 3401 (1995).

[29]   T. Bellini, V. Degiorgio, F. Mantegazza, F. A. Marsan, C. Scarneccia, *J. Chem. Phys.* **103**, 8228 (1995).

[30]   T. Palberg, W, Härtl, U. Wittig, H. Versmold, M. Würth, E. Simnacher: *J. Phys. Chem.* **96**, 8180 (1992).

[31]   M. Evers, N. Garbow, D. Hessinger, T. Palberg: *Phys. Rev. E* **57**, 6776 (1998).

[32]   T. Palberg: *J. Phys. Condens. Matter* **11**, 323 (1999).

[33]   J. E. Seebergh, J.C.Berg: *Colloids and Surfaces A* **100**,139 (1995).

[34]   C. F. Zukoski, D. A. Saville: *J. Colloid Interface Sci.* **114**, 32 and 45(1986).

[35]   D. R. Lide (Ed.): *Handbook of Chemistry and Physics*, CRC, Boca Raton [75]1994.

[36]   T. Okubo: *Ber. Bunsenges. Phys. Chem.* **91**, 1064 (1987).

[37]   K. Ito, N. Ise, T. Okubo: *J. Phys. Chem.* **82**, 5732 (1985).

[38]   S. Bucci, C. Fagotti, V. Degiorgio, R. Piazza: *Langmuir* **7**, 824 (1991).

[39]   J.Yamanaka, Y. Hayashi, N. Ise, T. Yamaguchi: *Phys. Rev. E* **55**, 3028 (1997).

[40]   T. Palberg, M. Evers, N. Garbow, D. Hessinger in S. C. Müller, J. Parisi, W. Zimmermann: *Transport and Structure in Biophysical and Chemical Phenomena*, Springer 1999, (in press).

[41]   L. Belloni: *Colloids and Surf. A* **140**, 227 (1998).